\documentclass[a4paper]{jpconf}
\usepackage{graphicx}
\usepackage{iopams}
\begin{document}
\title{Photoproduction of $\rho^0$ in ultra--peripheral nuclear collisions at ALICE}

\author{Kyrre Skjerdal, for the ALICE collaboration}

\address{Department of Physics and Technology, University of Bergen, P.O. Box 7803, 5020 Bergen, Norway}

\ead{kyrre.skjerdal@cern.ch}

\begin{abstract}
Photoproduction of $\rho^0$ mesons in ultra-peripheral Pb+Pb collisions has been
studied by the ALICE Collaboration at the CERN LHC. The strong photon
flux associated with relativistic charged nuclei leads to a very large cross
section for exclusive photoproduction of $\rho^0$ meson in interactions of the type
$Pb + Pb \rightarrow Pb + Pb + \rho^0$. For a $\rho^0$ produced at mid-rapidity at the LHC, the
photon-nucleus center of mass energy is higher than in
any previous experiment.

The ALICE detector is a general purpose detector dedicated to study heavy--ion collisions. ALICE has excellent performance in the low $p_T$ region, and can
reconstruct charged particle tracks with 0.1 GeV/c $\leq p_T \leq 100$ GeV/c.
In this analysis all tracks were required to be within ALICE's central barrel. Analysis of data from the first heavy ion run at the LHC in 2010 will be discussed in this paper. 




\end{abstract}

\section{Introduction to ultra--peripheral collisions}
\begin{figure}[h]
\includegraphics[width=18pc]{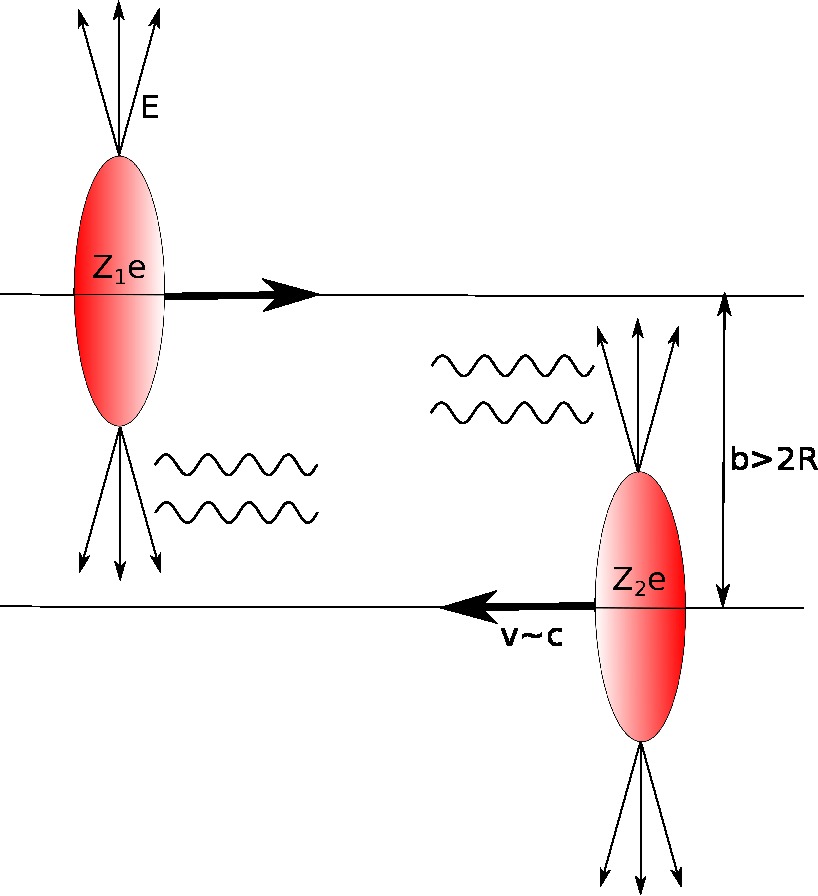}\hspace{2pc}%
\begin{minipage}[b]{18pc}\caption{\label{fig:upc1}The impact parameter, $b$, is larger than two times the radius, $R$, of the colliding hadrons. The photon flux from each nucleus is proportional to $Z^2$.}
\end{minipage}
\end{figure}

Ultra--peripheral collisions are collisions between hadrons, they can be protons or nuclei, where they geometrically miss each other. This implies that the impact parameter is larger than two times the radii of the colliding hadrons, see Figure \ref{fig:upc1}. For light vector meson production in Pb--Pb collisions at LHC energy, the median impact parameter is usually in the range 200 -- 300 fm. 
Because such impact parameters are much larger than the short range of the strong interaction, the interactions will be mediated by the electromagnetic field. The electromagnetic field of a moving charged particle can be treated as a flux of virtual photons. This model is referred to as the Weiz\"acker--Williams method. The intensity of the electromagnetic field, and therefore the number of virtual photons in the cloud is proportional to $Z^2$, where $Z$ is the charge of the particle \cite{baltz}. 




One can divide ultra-peripheral collisions into 
two categories: two-photon and photonuclear interactions. The photonuclear 
interactions can be further subdivided into coherent and incoherent 
interactions. In coherent photonuclear interactions, the photon interacts
with the whole target nucleus coherently. In most cases both nuclei will remain intact after the interaction.    For photonuclear interactions a photon from one nucleus is interacting with the other nucleus and, because the photon has spin $J^P = 1^-$, a vector meson can be produced. The transverse momentum of the resulting particles will be determined by the nuclear form factor, the average transverse momentum will be $\langle p_T \rangle \sim 60$ MeV/c.

In an incoherent interaction a photon from one nucleus interacts with a single nucleon in the target nucleus. This will cause the target nucleus to break up. The transverse momentum is higher in incoherent interactions than in coherent interactions, $\langle p_T \rangle \sim 400$ MeV/c for the $\rho^0$ \cite{starlight}. 

In two--photon interactions one photon from each of the two colliding nuclei collides and produces a fermion pair. An example is given by $Pb+Pb \rightarrow Pb+Pb + e^{+}e^-$, where the interaction is $\gamma+\gamma \rightarrow e^+e^{-}$.

For vector meson production in photonuclear coherent interactions the event will contain only two tracks, and the detector will otherwise be empty. There will be one positive and one negative track, and the transverse momentum of the track pair will be $\sum p_T \lesssim 100$ MeV/c \cite{starlight, star}. 

\section{Model predictions}


The center--of--mass energy per nucleon in Pb--Pb collisions at the LHC in the 2010 and 2011 Pb--Pb runs was $\sqrt{s_{NN}} = 2.76$ TeV. This corresponds to a Lorentz factor of $\gamma_{L} = 1470$ for each beam in the center of mass system. The $\rho^0$ is a broad resonance with mass $M_{\rho} = 775$ MeV/c$^2$ and width $\Gamma_{\rho} = 149$ MeV/c$^2$.  At mid--rapidity, $y=0$, the mass 775 MeV/c$^2$ corresponds to a mean photon energy of $E_{\gamma} = 775\textrm{ MeV}/2 = 338$ MeV. The equivalent $\gamma$--proton center--of--mass energy is then 
\begin{equation}
W_{\gamma p} = \sqrt{4E_{\gamma}m_{p}\gamma_{L}} = 43.2 \textrm{ GeV}. 
\end{equation}
This is three times higher than at RHIC, and higher than in any fixed--target experiment.

Three different models predict the cross section for $\rho^0$ production at LHC energies. 
The model by Frankfurt, Strikman, and Zhalov (GGM (\textit{Gribov--Glauber Model})) \cite{ggm1, ggm2} uses a generalized vector dominance model in the Gribov--Glauber approach. It includes non--diagonal transitions, where the photon fluctuates to a $\rho'$, but appear as a $\rho^0$ after scattering off the target nucleus. The cross section $\sigma(\rho + \textrm{nucleon})$ from the Donnachie--Landshoff  model, which is in agreement with HERA and lower energy data, is used as input for the photonuclear calculation. 

\begin{figure}[h]
    \includegraphics[width=18pc]{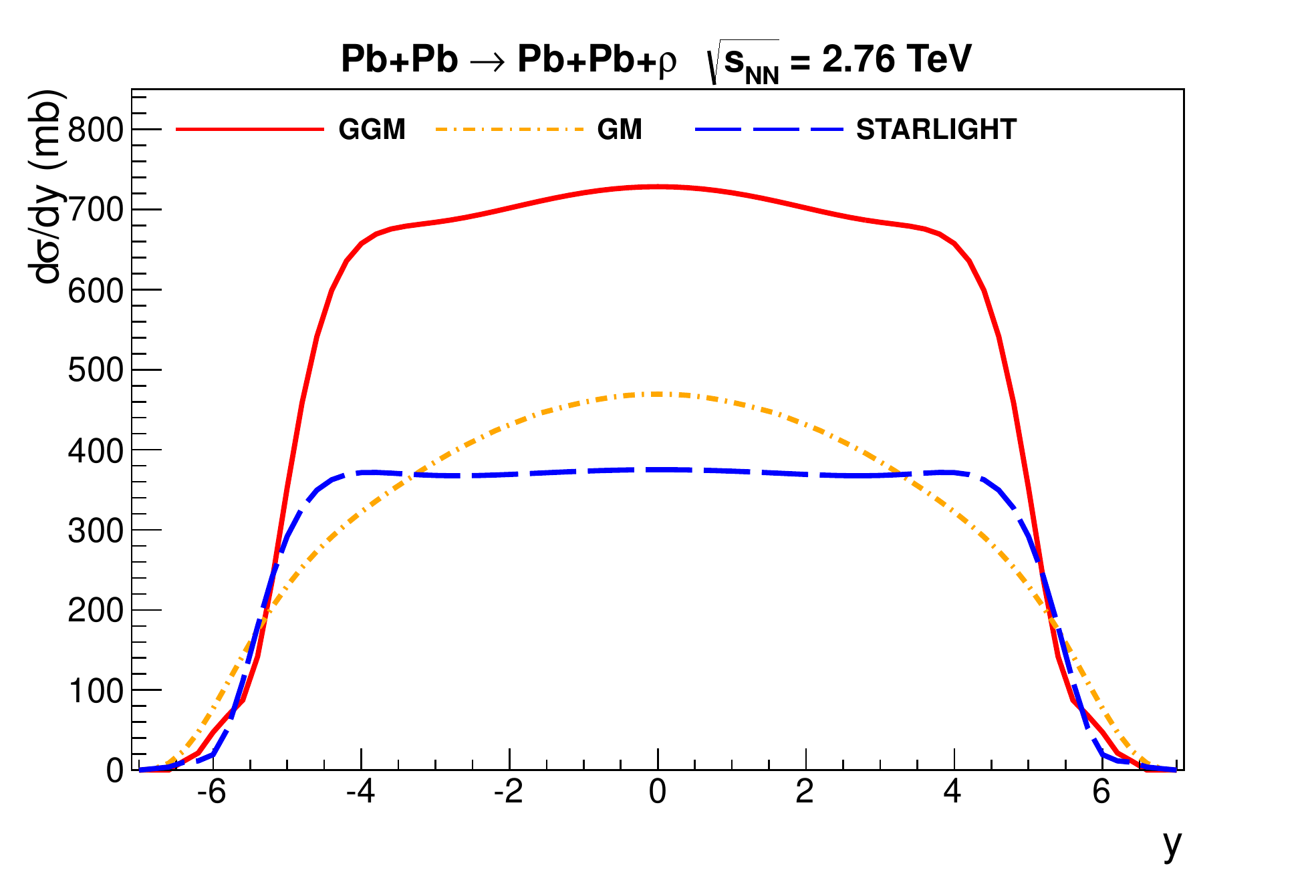}
    \begin{minipage}[b]{18pc}\caption{\label{predictions} Model predictions for $\rho^0$ photoproduction at central rapidities ($y=0$), for the models GGM \cite{ggm1, ggm2}, shown with a full (\full) red line, GM \cite{gm}, shown with a yellow chained (\chain) line and Starlight \cite{starlight, starlighturl}, shown with a blue dashed (\dashed) line. (Color online.)}
    \end{minipage}
\end{figure}

The model developed by Gon\c{c}alves and Machado (GM) \cite{gm} is based on the color dipole model in combination with saturation from a Color Glass Condensate approach. Starlight \cite{starlight, starlighturl} is a Monte Carlo event generator developed by Klein and Nystrand. Experimental data for $\gamma + p \rightarrow \rho^0 + p$ is used in combination with a Glauber model, neglecting the elastic scattering. The predictions from the different models can be seen in Figure \ref{predictions}.



\section{The ALICE detector} 
\begin{figure}[h]
  \begin{center}
  \includegraphics[width=0.7\textwidth]{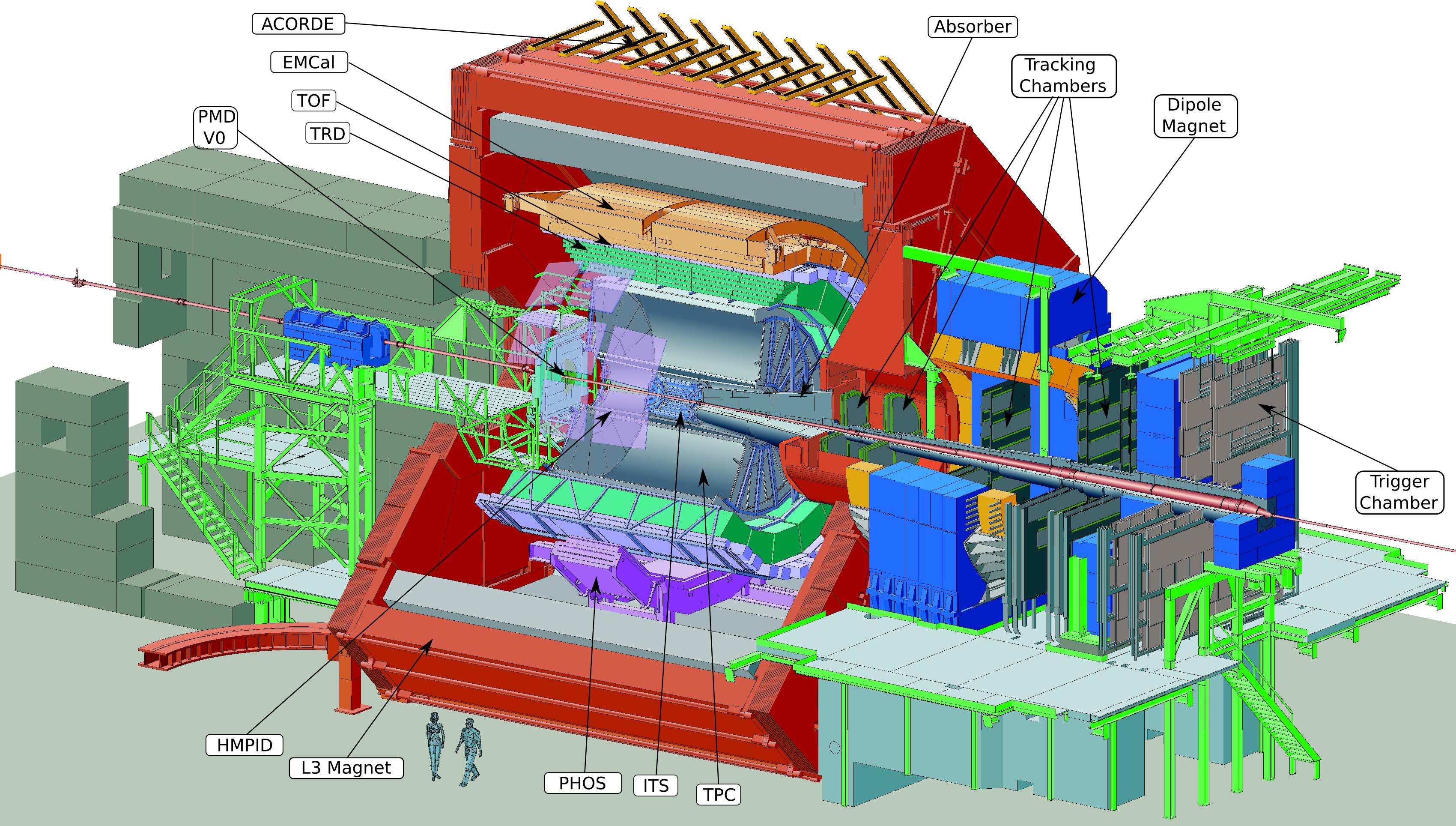}
  \caption{\label{fig:alice}The ALICE detector}
  \end{center}
\end{figure}

The ALICE detector (Figure \ref{fig:alice}) is a general purpose detector at the CERN LHC \cite{alice}. Its main goal is to study ultra--relativistic heavy--ion collisions. It consists of a central barrel, a forward muon arm, and some other smaller forward detectors. The central barrel has a pseudorapidity acceptance of $|\eta| < 0.9$ and a transverse momentum acceptance of $p_T > 100$ MeV/c. 

For this analysis mainly the central barrel is used. The Inner Tracking System (ITS) and the Time Projection Chamber (TPC) are used for tracking and particle identification. The ITS consists of six layers of silicon detectors; the two innermost layers are silicon pixel detectors (SPD), the two next are silicon drift detectors (SDD), and the two outermost layers are silicon strip detectors (SSD). The ALICE TPC, a cyllindrical gaseous detector with a diameter of 500~cm and a length of 510~cm, is the main tracking device in ALICE. The 557,568 readout pads can provide up to 159 ionization samples for track reconstruction, which can be used to calculate the energy loss (d$E$/d$x$) of the track. The dE/dx of a track is calculated as the truncated mean of the d$E$/d$x$ of the clusters associated with the track. The truncated mean is used in order to reduce the fluctuations in cluster energies resulting from the Landau tail \cite{alicepid}.
For triggering the SPD, the Time--of--Flight detector (TOF) and the VZERO counters are used. The TOF detector, which surrounds the TPC, is composed of multigap resistive plate chambers which provide an intrinsic resolution of approximately 80 ps. The VZERO counters, on each side of the interaction point, consists of 32 tiles of scintilators. They are in this analysis used to define rapidity gaps. The VZERO--A has an acceptance of $2.8 \leq \eta \leq 5.1$, and the VZERO--C has an acceptance of $-3.7 \leq \eta \leq -1.7$. 
To count neutrons from nuclear break up, the Zero Degree Calorimeters ({\it ZDC}), which are hadronic calorimeters, located $\pm 116$ meters on each side of the interaction point, are used.   

\section{Analysis of 2010 Pb--Pb data}
\subsection{Data sets and cuts}
For this analysis data recorded during the 2010 Pb--Pb run, at center--of mass energy $\sqrt{s_{NN}} = 2.76$ TeV, are used. In the first part of the run, when the luminosity was low, a trigger which required at least two hits in the TOF detector was used. Later in the run, when the luminosity was higher, a tighter trigger definition was needed. A trigger which required at least two hits in the TOF detector, at least two hits in the Si--pixel detectors and no activity in the VZERO counters was implemented for the last part of the run. 

\begin{figure}[h]
  \begin{minipage}{18pc}
    \includegraphics[width=18pc]{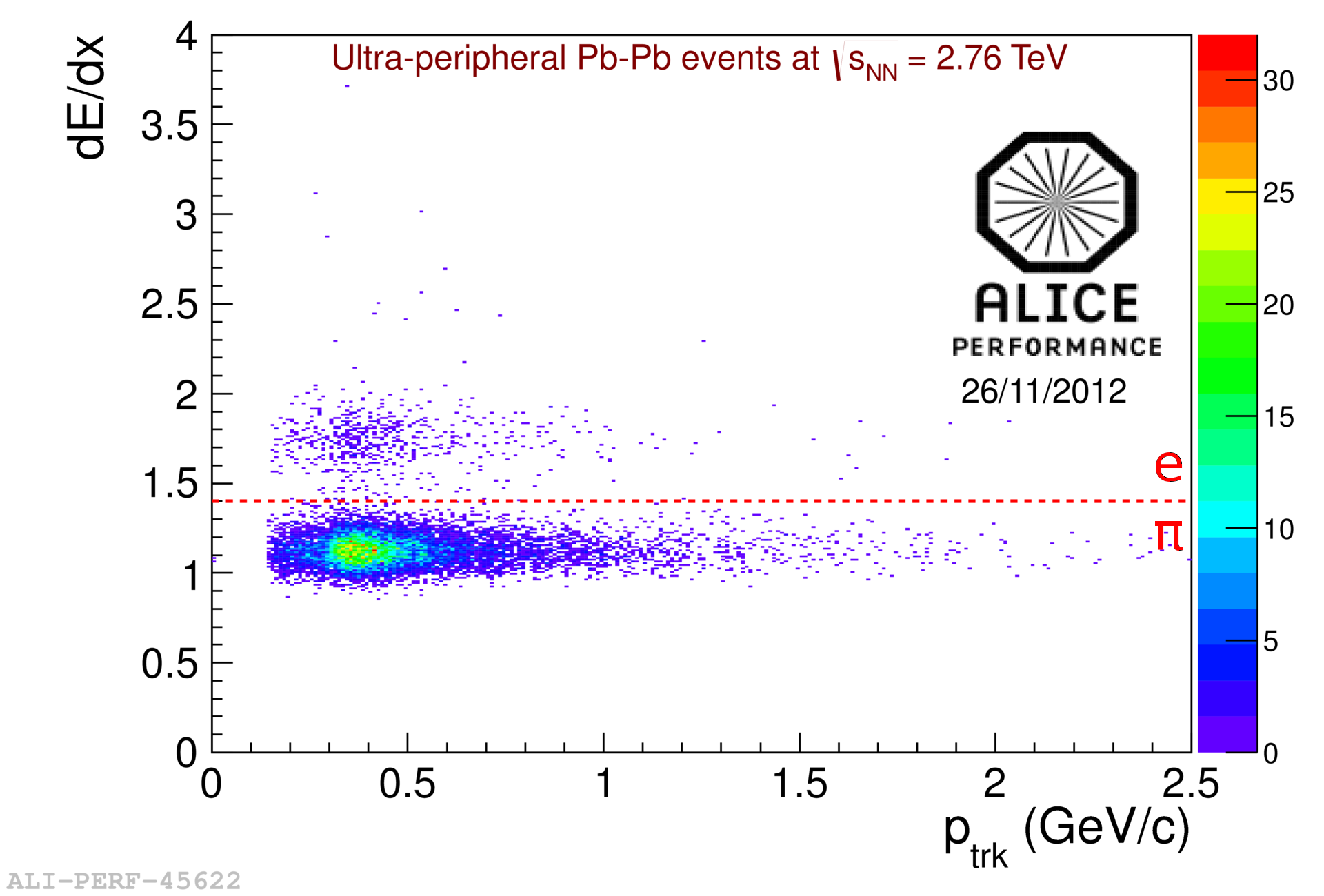}
    \caption{\label{fig:dedxvsp}Energy loss in the TPC plotted versus the momentum of the track, for $\rho^0$ candidates \\}
  \end{minipage}\hspace{2pc}%
  \begin{minipage}{18pc}
    \includegraphics[width=18pc]{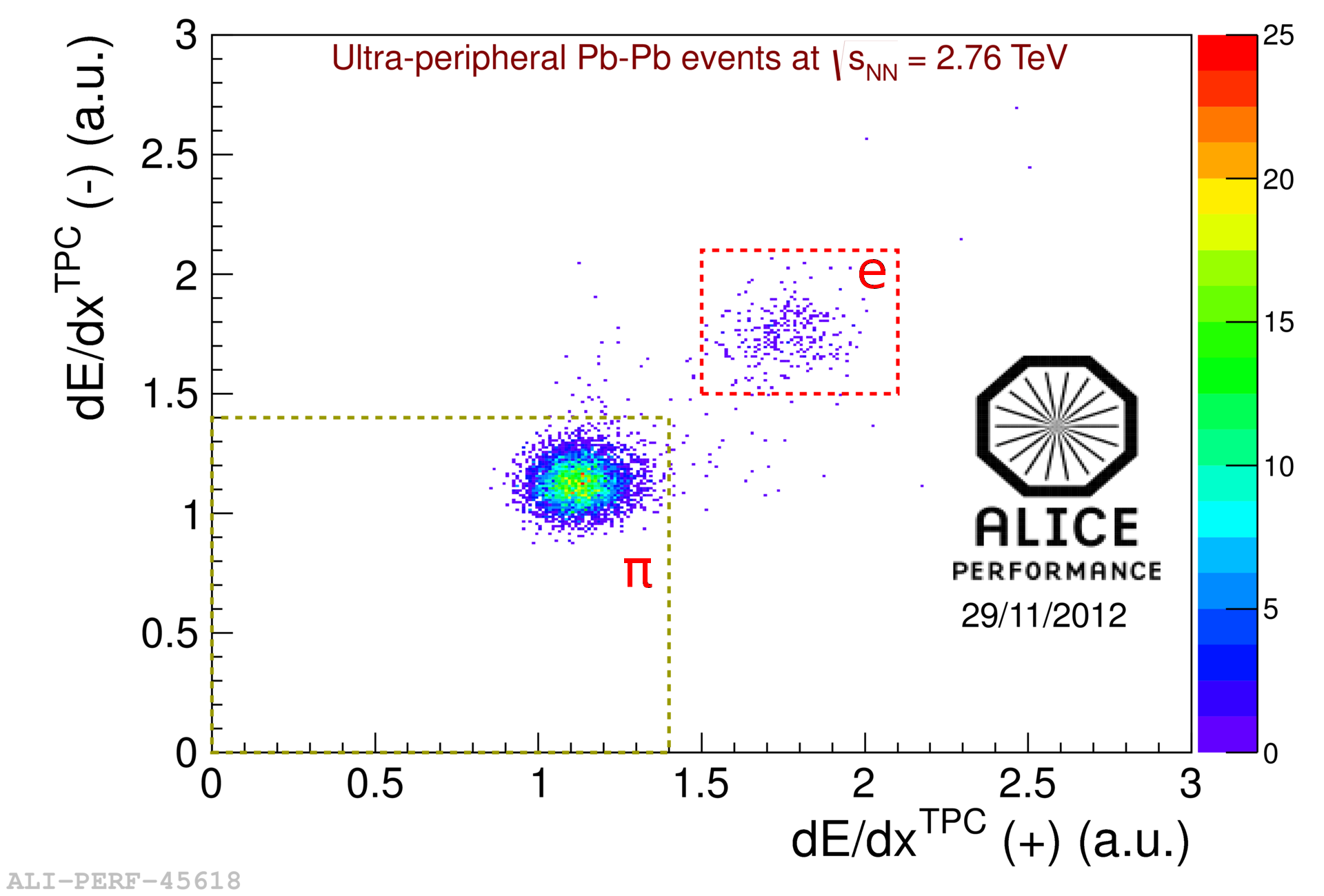}
    \caption{\label{fig:dedx1vsdedx2}Energy loss of the positively charged particles vs. the energy loss of the negatively charged particles in the TPC, for $\rho^0$ candidates.}
  \end{minipage} 
\end{figure}

Based on the characheristics of UPC events, a set of cuts are applied to the data at analysis level.
The events are required to satisfy one of the two UPC triggers at hardware level. The event must have a primary vertex, where the z--position (along the beam axis) is within 10 cm from the center. There must be exactly two accepted ITS+TPC tracks, and the VZERO detectors should be empty. The two tracks are required to produce ionization in the TPC consistent with pions (Figure \ref{fig:dedxvsp} and \ref{fig:dedx1vsdedx2}). The rapidity of the mother particle should be $|y_{pair}| < 0.5$ and the transverse momentum required to be below $p_T^{pair} < 150$ MeV/c, to get the coherent events. In the end it was required that the tracks have opposite charge, and the like sign background (less than 2\%) was subtracted. 
The cuts, and the number of events surviving each of them, are listed in Table \ref{tab:cutsc0om2} and Table \ref{tab:cutsccup2}. 
\begin{table}[h]
  \begin{minipage}{18pc}
  \caption{\label{tab:cutsc0om2}Number of events surviving the cuts for the TOF--trigger.} 
  \begin{center}
    \begin{tabular}{ll}
      \br
      Cut&Events left\\
      \mr
      Triggered events&1,332,041\\
      Primary vertex&850,409\\
      Two accepted tracks&47,978\\
      $|V_z|<10$ cm&43,413\\
      VZERO veto&8,848\\
      PID cut&7,588\\
      $|y|<0.5$&5,887\\
      $p_T^{pair}<150$ MeV/c&2,749\\
      Unlike sign pairs&2,699\\
      Like sign pairs&50\\
      \br
    \end{tabular}
  \end{center}
  \end{minipage}\hspace{2pc}
  \begin{minipage}{18pc}
    \caption{\label{tab:cutsccup2}Number of events surviving the cuts for the SPD+TOF+VZERO vetoed--trigger.} 
  \begin{center}
    \begin{tabular}{ll}
      \br
      Cut&Events left\\
      \mr
      Triggered events&121,487\\
      Primary vertex&103,480\\
      Two accepted tracks&26,217\\
      $|V_z|<10$ cm&24,020\\
      VZERO veto&17,567\\
      PID cut&15,377\\
      $|y|<0.5$&11,928\\
      $p_T^{pair}<150$ MeV/c&6,195\\
      Unlike sign pairs&6,101\\
      Like sign pairs&94\\
      \br
    \end{tabular}
  \end{center}
  \end{minipage}
\end{table}

\subsection{Acc $\times$ Eff correction}
To correct for acceptance and efficiency a flat invariant mass $\pi^+\pi^-$ simulation is used. The flat simulation is chosen because the shape of the $\rho^0$ peak would give low statistics for the correction at the tails of the distribution. It is assumed that the $\pi^+\pi^-$ pairs are emitted from a transversely polarized parent, as expected for coherently produced $\rho^0$'s. The generated $\pi^+\pi^-$ events are processed by the ALICE simulation and reconstruction framework using the Geant transport  to simulate the detector response. These simulated events are then passed through the same analysis as is used on the data. The $(Acc\times Eff)$  is defined as the ratio of the number of reconstructed to generated selected events in the rapidity interval $|y| < 0.5$ and $p_T < 150$ MeV/c as a function of invariant mass in the range $2m_{\pi} < M_{inv} < 1.5$ GeV/c$^2$. 
\subsection{Fitting the invariant mass distribution}
The invariant mass distribution is corrected for $(Acc\times Eff)$  and fitted with a Breit--Wigner function with continuum correction (Equation \ref{eq:breitwigner}). 
\begin{equation}
  \frac{d\sigma}{dM_{\pi\pi}} = \left|A\frac{\sqrt{M_{\pi\pi}M_{\rho^0}\Gamma(M_{\pi\pi})}}{M^2_{\pi\pi}-M^2_{\rho^0}+iM_{\rho^0}\Gamma(M_{\pi\pi})}+B\right|^2
  \label{eq:breitwigner}
\end{equation}
where
\begin{equation}
\Gamma(M_{\pi\pi}) = \Gamma_0 \cdot (M_{\rho^0}/M_{\pi\pi}) \times [(M_{\pi\pi}^2 - 4m_{\pi}^2)/(M_{\rho^0}^2 - 4m_{\pi}^2)]^{3/2}
\end{equation}
is the momentum dependent width of the $\rho$ meson, A is the amplitude of the Breit--Wigner function and B is the amplitude of the direct non--resonant $\pi^+\pi^-$ production.This function has previously been used by the STAR \cite{star} and H1 \cite{h1} collaborations. 

\begin{figure}[h]
  \begin{minipage}{18pc}
    \includegraphics[width=18pc]{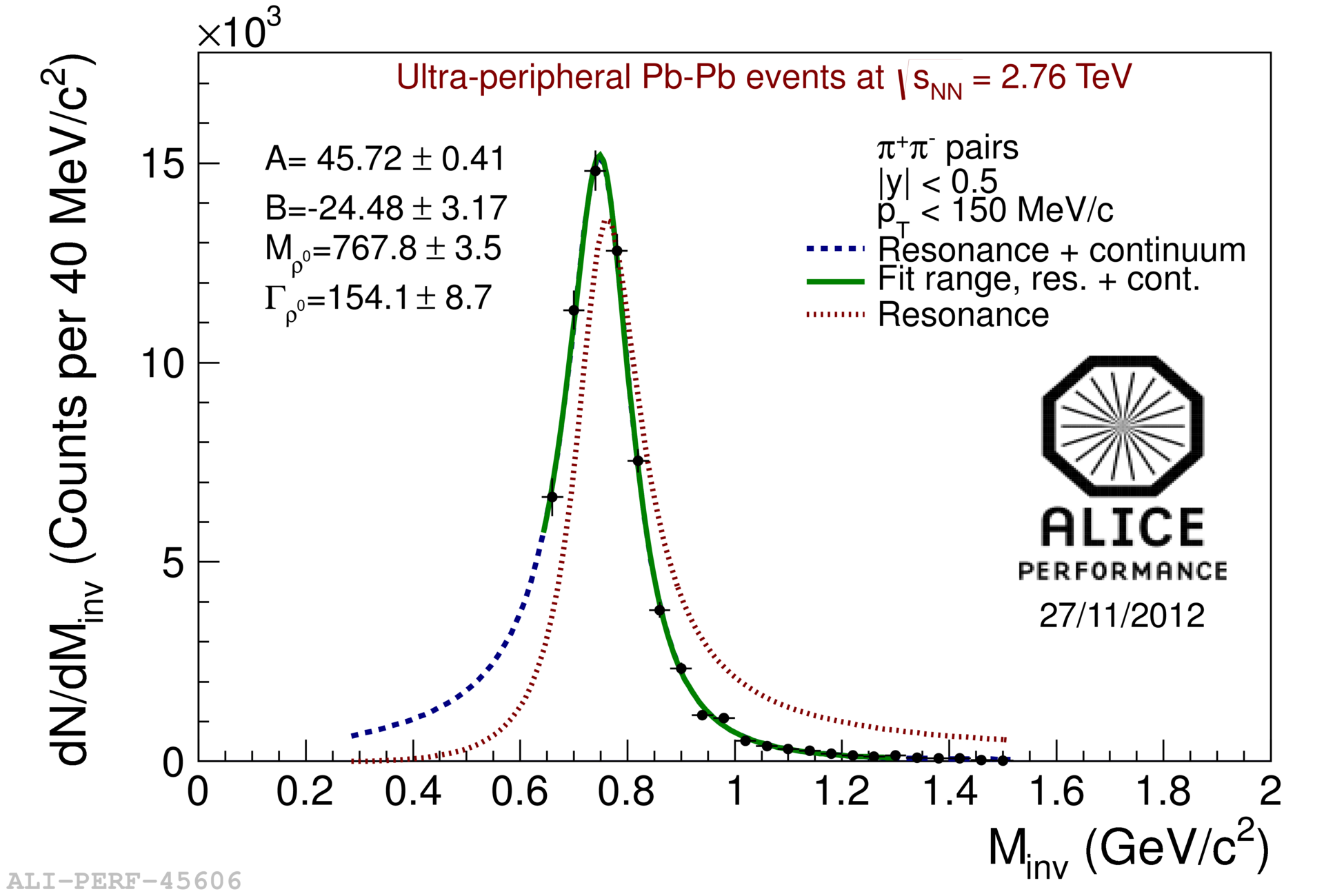}
  \end{minipage}\hspace{2pc}%
  \begin{minipage}{18pc}
    \includegraphics[width=18pc]{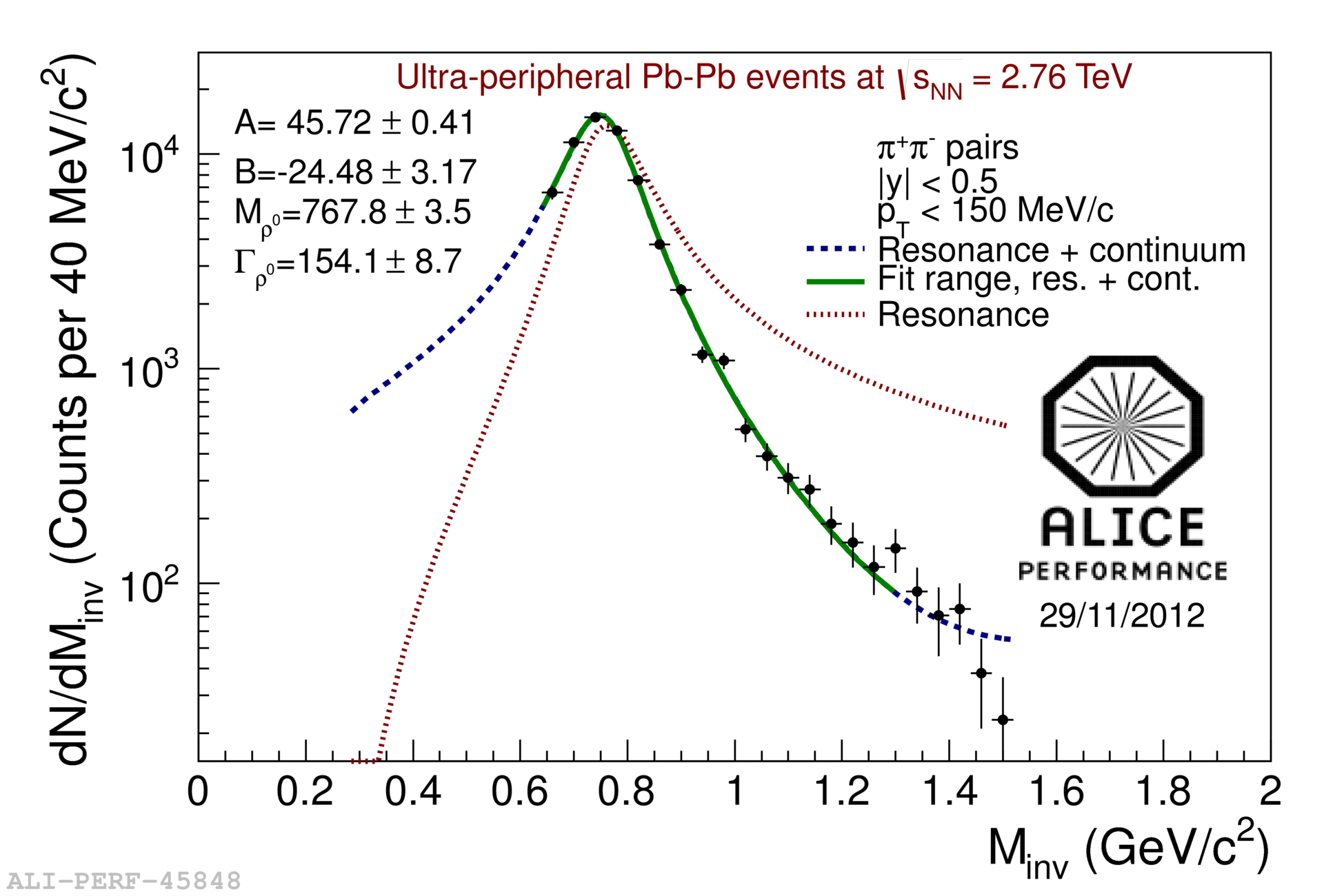}
  \end{minipage} 
  \caption{\label{fig:bwfit}The invariant mass distribution fitted with a Breit--Wigner function with continuum correction. Linear (left) and logartimic (right) scale. The data points are marked with a full circle (\fullcircle), the blue dashed line (\dashed) is the resonance plus the continuum, the full green line (\full) is the resonance plus the continuum in the range used to make the fit and the red dotted line (\dotted) is only the resonance. Statistical errors are shown. (Color online.)}

\end{figure}

The fitted invariant mass distribution is shown in linear scale in the left panel of Figure \ref{fig:bwfit}, and in logaritmic scale in the right panel. The mass of $M_{\rho^0} = 767.8 \pm 3.5$ MeV/c$^2$ and the width of $\Gamma_{\rho^0} = 154.1 \pm 8.7$ MeV/c$^2$, are compatible with the PDG values of $M = 775.49 \pm 0.34$ MeV/c$^2$ for the mass, and $\Gamma = 149.1 \pm 0.8$ MeV/c$^2$ for the width.

\subsection{Subtraction of incoherent contribution}
The transverse momentum cut $p_T < 150$ MeV/c will leave mostly the coherent events, but also some incoherent events will remain. To account for this, one has to find the fraction of incoherent events with $p_T < 150$ MeV/c. To do this the Starlight \cite{starlight, starlighturl} particle generator is used. The detector response is also simulated, and the simulated particles are reconstructed with the ALICE analysis framework \cite{aliroot}.  


\begin{figure}
  \begin{center}
    \includegraphics[width=0.7\textwidth]{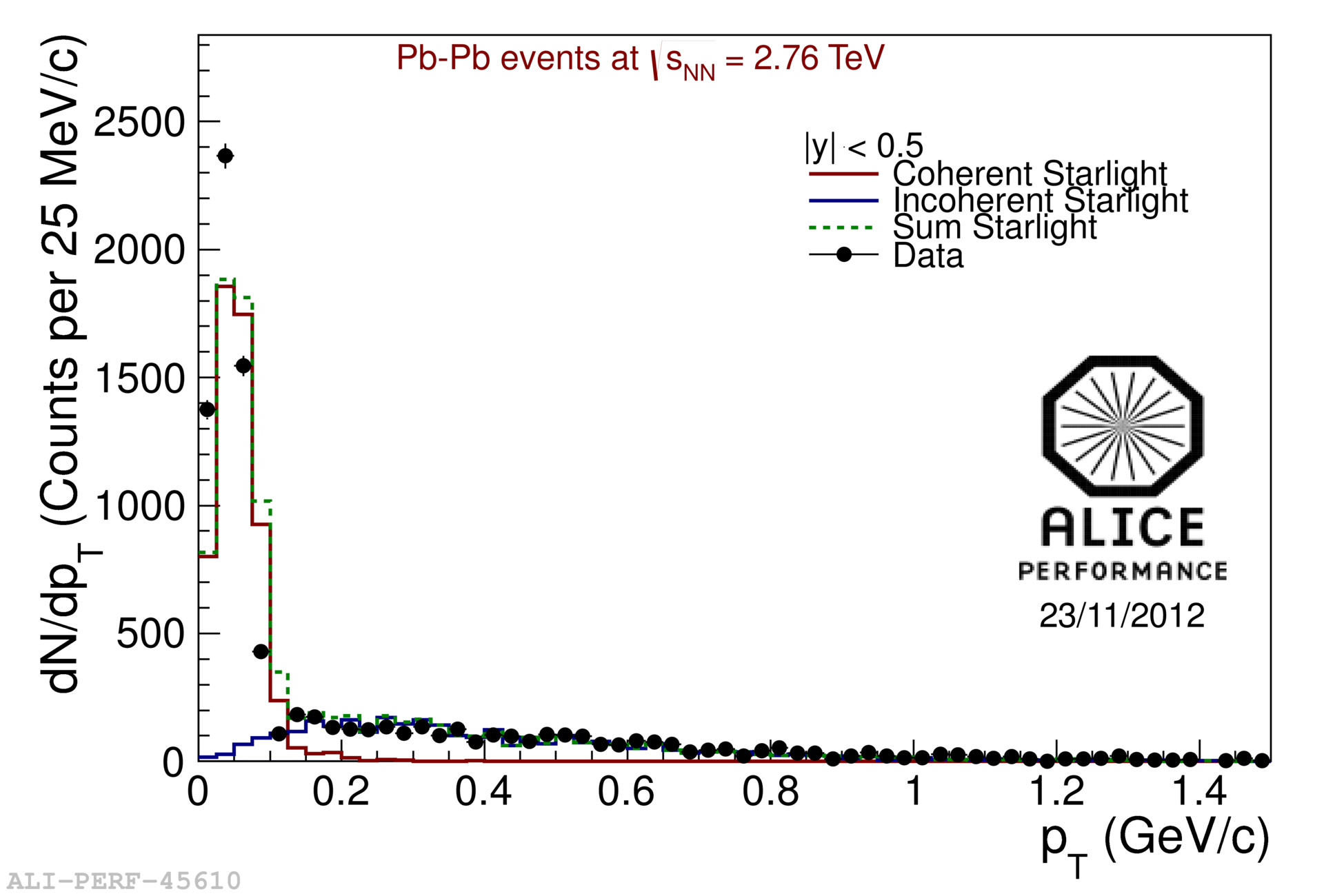}
    \caption{\label{fig:pt} The transverse momentum distribution compared to Starlight predictions for coherent and incoherent photoproduced $\rho^0$'s. The data points are marked with a full circle (\fullcircle), the simulated coherent production is marked with a red full line (\full), the simulated incoherent production is marked with a full blue line (\full), and the sum of the incoherent and coherent simulation is marked with a dashed green line (\dashed). (Color online.)}
  \end{center}
\end{figure}

The data points and the simulated distributions are shown in Figure \ref{fig:pt}. A coherent and an incoherent simulated sample was generated. The two samples were scaled to fit the data. The contribution from incoherent events under the coherent peak ($p_T < 150$ MeV/c) is found to be $\sim 7\%$. 
The coherent peak appears to be slightly narrower in data than in the simulation. This could be a detector effect or related to the implementation of the nuclear form factor in the Monte Carlo. 
\subsection{Nuclear break up}

\begin{figure}[h]
  \begin{minipage}{18pc}
    \includegraphics[width=18pc]{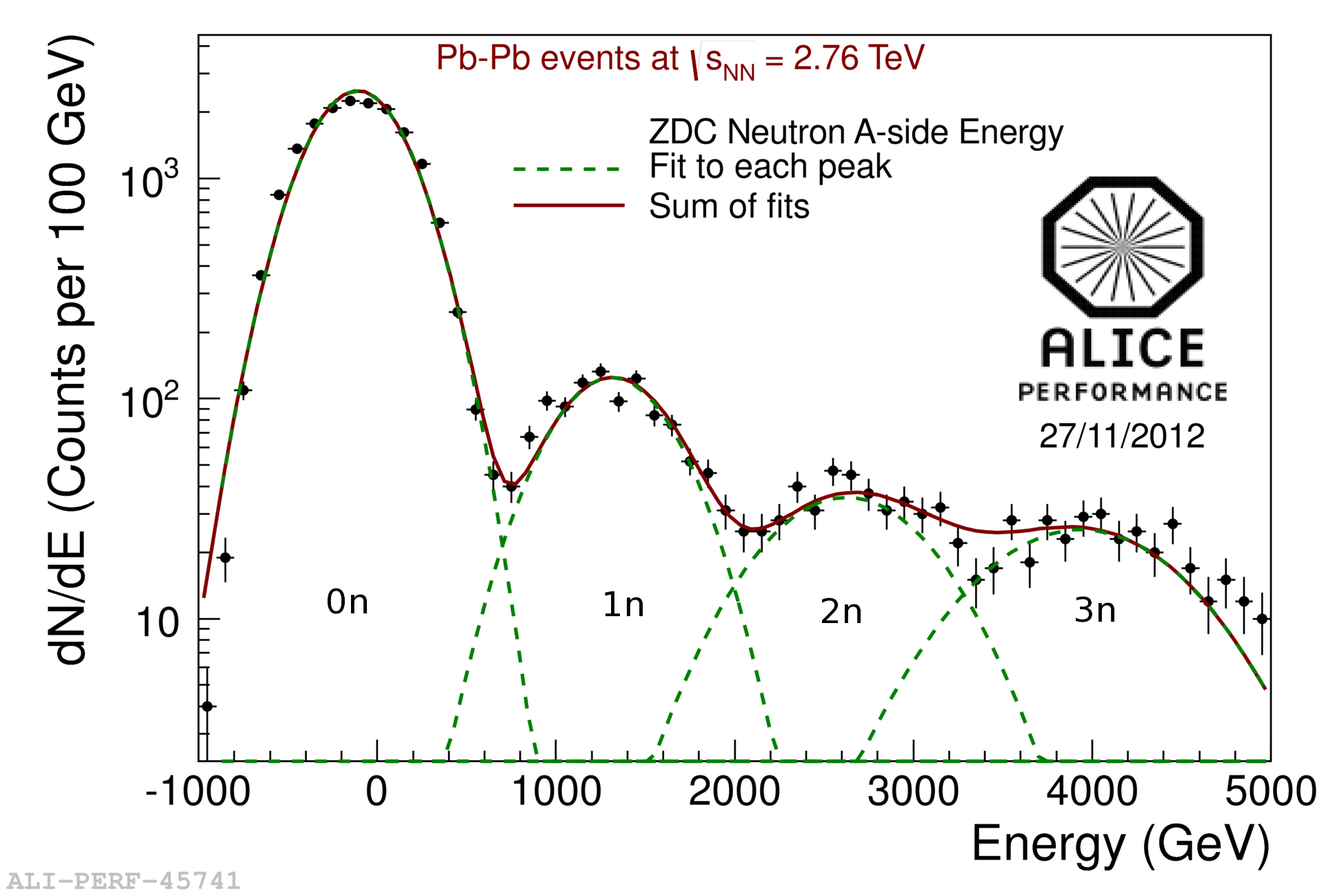}
  \end{minipage}\hspace{2pc}%
  \begin{minipage}{18pc}
    \includegraphics[width=18pc]{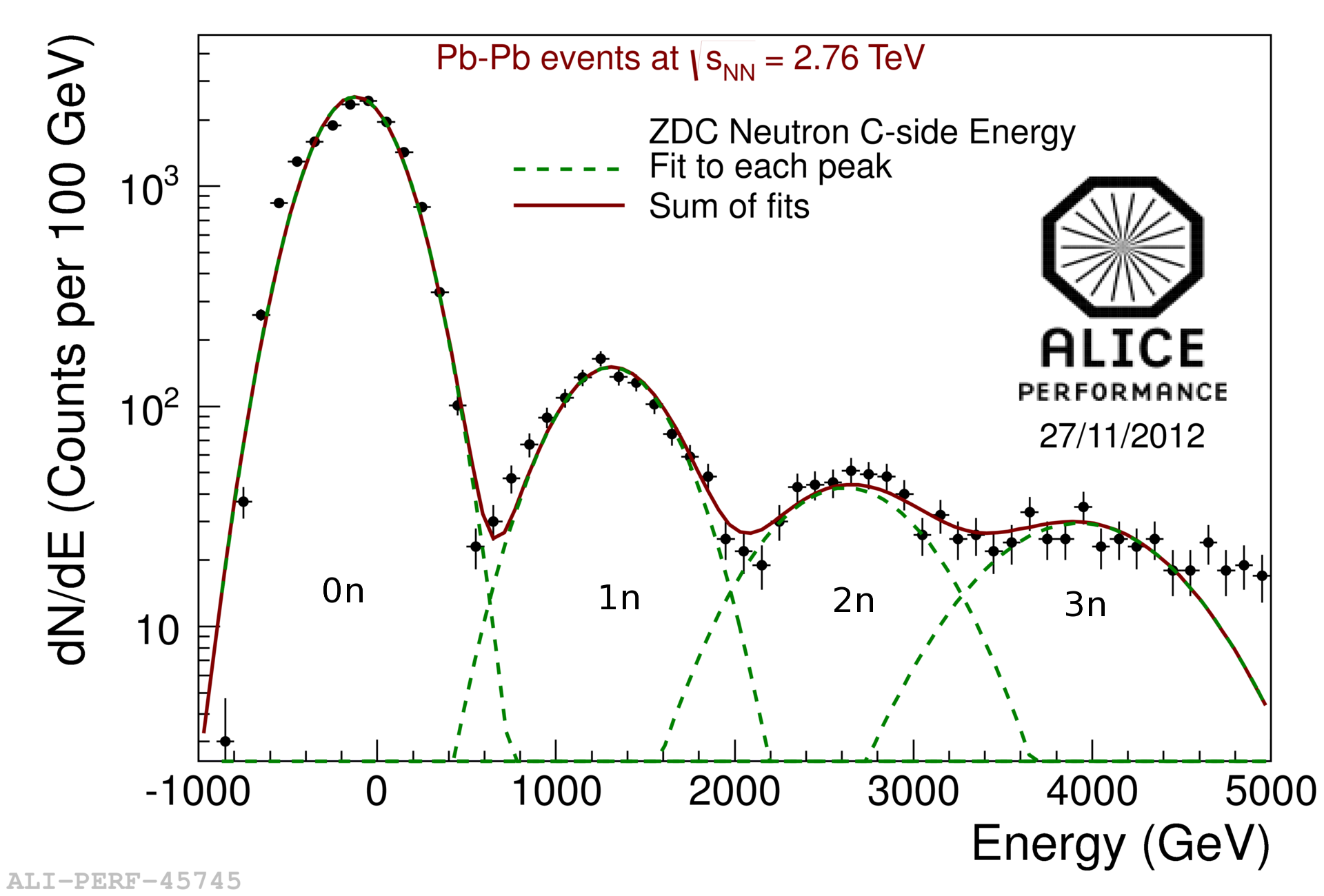}
  \end{minipage} 
\caption{\label{fig:zdcenergy}Energy deposited in the ZDC on the A--side (left) and the C--side (right). Each peak is fitted with a Gaussian. The first peak represent zero neutrons, the second peak one neutron and so on. Because a pedestal value is subtracted from the signal, the energy goes below zero. The data points are marked with a full circle (\fullcircle), the fit to each peak each marked with a green dashed line (\dashed), and the sum of the fits is marked with a red full line (\full). (Color online.)}
\end{figure}

Exchange of additional photons may lead to coherent vector meson production in coincidence with nuclear break up \cite{baltz2002}. To count the number of neutrons emitted from the nuclei, the ZDC neutron counters, located at 116 meters on each side of the interaction point, are used. Figure \ref{fig:zdcenergy} shows the energy deposited in each of the ZDCs, for triggered events with two accepted tracks, and the z--position of the primary vertex within 10 cm from the center. The peaks corresponding to zero and one neutron in the ZDCs can be clearly identified. 
The peak centered around $E = 1380$ GeV, which is the beam energy per nucleon, represents one neutron detected. 

In the final selection one distinguishes between the two cases of zero neutrons and one or more neutrons in the ZDC. The separation between no neutrons and one or more neutrons is defined from the minimum between the first two peaks in Figure \ref{fig:zdcenergy}. As expected, the energy at this minimum corresponds to half the beam energy per nucleon.
This distinction will make it possible to meassure the cross section for photoproduction of $\rho^0$ with and without nuclear break up seperately, and this can be compared with model predictions \cite{baltz2002}.
 
\section{Conclusions and outlook}
The exclusive photonuclear production of $\rho^0$ has been studied by the ALICE collaboration. The analysis cuts developed for ultra--peripheral collisions allow a clear separation of the signal, with a background contribution estimated from the like-sign contribution of less than 2\% (cf. Tables \ref{tab:cutsc0om2} and \ref{tab:cutsccup2}). The invariant mass and transverse momentum distributions generally agree with the expectations, although the coherent peak in the $p_T$ distribution appears slightly narrower in data. 
The measurement of the cross section at mid-rapidity will allow a better understanding of $\rho^0$ photoproduction. 
\\
\section*{References}

\end{document}